\documentclass[%
 reprint,amsmath,amssymb,aps,showpacs]{revtex4-1}

\usepackage{epsfig}
\usepackage{graphicx}
\usepackage{amssymb}

\newcommand {\ket}[1] {|{#1}\rangle}

\begin{document}


\title{Functional Quantum Computing: An Optical Approach}


\author{Timothy M. Rambo}
\email{timothyrambo2017@u.northwestern.edu}
\author{Joseph B. Altepeter}
\altaffiliation{JBA is currently affiliated with Berberian \& Company, LLC, but his authorship here reflects work performed before joining that company.}
\author{Prem Kumar}
\affiliation{Center For Photonic Communication and Computing, EECS Department, Northwestern University,\\ 2145 Sheridan Road, Evanston, IL 60208-3118, USA}%

\author{G. Mauro D'Ariano}
\affiliation{QUIT Group, Dipartmento di Fisica, Universit\`{a} di Pavia, and INFN, via Bassi 6, 27100 Pavia Italy}

\date{\today}

\date{\today}

\begin{abstract}
Recent theoretical investigations treat quantum computations as functions, quantum processes which operate on other quantum processes, rather than circuits. Much attention has been given to the $N\!-switch$ function which takes $N$ black box quantum operators as input, coherently permutes their ordering, and applies the result to a target quantum state. This is something which cannot be equivalently done using a quantum circuit. Here, we propose an all-optical system design which implements coherent operator permutation for an arbitrary number of input operators.
\end{abstract}
\pacs{03.67.-a,42.50.-p} 

\maketitle

\section{Introduction}
\paragraph*{} Quantum computers hold the promise of dramatic speed increases over their classical analogues for certain types of problems, such as search\cite{GroverSearch}, factoring\cite{ShorFactor}, and many others (see \cite{MikeAndIke} for a summary of the topic). Specific quantum algorithms are typically described using quantum circuits, which represent quantum computations at a low level in a qubit-wise step-by-step manner, similar to representing a classical computation using machine code. Describing computations in this manner, while accurate, can be cumbersome and the resulting circuit diagrams can be difficult to understand intuitively. By developing a quantum analogue to the classical model of functional computing, it
may be possible to replace the quantum circuit description with functional formalism which is more intuitive and has different capabilities.

\paragraph*{}A classical machine code predefines a set of bit-wise operations to be applied in a fixed order for any input bit register values. However, the broad scope of a complex program such as an operating system or graphical user interface is difficult to comprehend when focusing on every transformation of every bit value. Complex computations are instead designed in the abstract as a collection of functions, black boxes which can be applied arbitrarily and return data and/or other functions.

\paragraph*{}

Investigations into quantum functions find that while physical systems are capable of arbitrarily controlling the use and ordering of black box unitaries, quantum circuits cannot represent this\cite{D'ArianoSwitch,D'ArianoProgrammableQuantumComputation,NoControlBlackBox,PhysicalBlackBoxControl,PhysicalBlackBoxControlTransmon,Modularity} without significantly increased resources. Though this doesn't mean that quantum functions will necessarily lead to more physically resource efficient experiments, the functions can provide different and more compact formalism for conceptualizing quantum information processes. This is best illustrated by the $N\!-switch$ operation\cite{D'ArianoSwitch,D'ArianoProgrammableQuantumComputation} which coherently permutes the orderings of $N$ black box operators based on the value of control qubits and applies them to a target qubit register. In addition to being an interesting example of the differences between quantum functions and quantum circuits, arbitrary control over causal orderings is a useful mechanism for investigating the fundamental role of causality in quantum systems\cite{QuantumCorrWithNoCausalOrder,QuantumGravityComputers,QuantumCausalityInformationInsghts,QuantumCausality} and enhancing the efficiency of certain quantum information processing tasks\cite{PerfectNoSignallingDiscrimmination,ComputationalAdvantage}. A recent experiment has implemented $2\!-switch$ using spatial modes to control operator ordering, but to our knowledge there have been no practical proposals for permuting more than $2$ operators. Here we present an all-optical, experimental, architecture for $N\!-switch$, which exploits ultrafast quantum
coherent switching devices \cite{HallSwitchPRL, HallSwitchNJP,
xbarswitch} to permute an \textit{arbitrary} number of quantum operators while only requiring a single physical device to implement each quantum operator. The reader is cautioned that although ``quantum switch" is sometimes used synonymously with $N\!-switch$, in this paper quantum switch refers to an optical device for switching the path of photonic signals.

\section{Functional Quantum Computing}

\paragraph*{} The original classical model of functional computing is
the \textit{$\lambda$-calculus} \cite{lambdacalc}, which treats functions and data
as the same type
of objects. In other words, functions can be input to and output from other functions along with bit values.
One attempt at developing a functional definition of quantum computation is the
theory of ``quantum combs'' \cite{D'ArianoCombs1,D'ArianoCombs2}. Quantum combs are essentially quantum circuits with sockets that black box quantum operators can be
``plugged into,'' similar to micro-chips on a circuit board. However, this constructs a fixed circuit, granting the user only classical control over the operators. Other noteworthy proposals for functional quantum computing are
\cite{VanTonder} and \cite{LCwithClassicalControl}. 

More recently, quantum functions have been described as circuits with coherently controlled, movable, wires \cite{D'ArianoProgrammableQuantumComputation}. Movable wires grant the ability to reconfigure the connections between quantum operators in a computation at run-time and coherent control can be used to create a superposition of different uses\textemdash quantum control over black box quantum operators. This also paints a physical picture of a device with programmable interconnections, the implementation of which is key to our design. Similar investigations into quantum controlled black-box operators have also been carried out by other groups\cite{NoControlBlackBox,PhysicalBlackBoxControl,PhysicalBlackBoxControlTransmon,Modularity}.

This new functional model is showcased by the $N\!-switch$ operation\cite{D'ArianoProgrammableQuantumComputation}, which has an elegant functional definition but a complex equivalent quantum circuit. This $N\!-switch$ operator takes as input: $N$ black box quantum operators, $U_0$,$U_1$,...,$U_{N-1}$, a register of control qubits, and a register of input qubits. It coherently orders the input quantum operators based on the value of the
control
qubits, creating a meta-operator---a superposition of quantum operators in (potentially) many
different orderings---that is then applied to the input
qubit register \cite{D'ArianoProgrammableQuantumComputation}. To illustrate the differences between the circuit-based and functional descriptions, consider the simplest case:
$2\!-switch$.

\begin{figure*}[htb]
\centerline{
\includegraphics{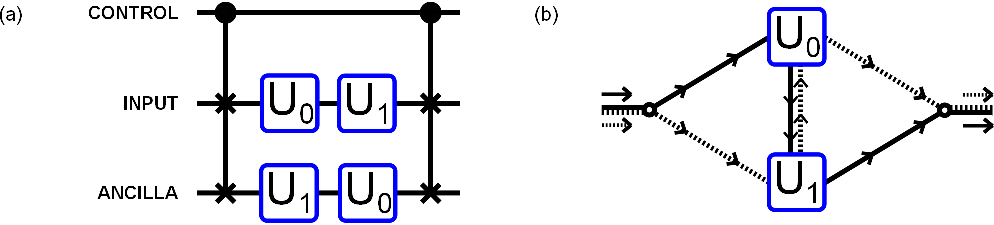}
}
\caption{\label{FIG:theoryQpermute} (a) A quantum circuit diagram which implements $2\!-switch$ using two copies of each operator. (b) A functional circuit diagram of $2\!-switch$ showing a superposition of two possible connections (solid and dashed) made by quantum controlled movable wires. Arrows illustrate the direction of information flow.}
\end{figure*}

The $2\!-switch$ operator takes the following inputs: two quantum operators $U_0$ and $U_1$, a control qubit $\ket{c} = \alpha\ket{0} + \beta\ket{1}$, and an input qubit $\ket{\psi}$. The $2\!-switch$ function (shown in eq. \ref{EQ:2switch} as $f_{2\!-switch}\left(\bullet\right)$ for clarity) outputs \begin{equation}\label{EQ:2switch}
\begin{split}                                                                                   f_{2\!-switch}\left(U_0,U_1,\ket{c},\ket{\psi}  \right) = &
\alpha\ket{0}\otimes
U_0 U_1 \ket{\psi} \\ &
+\beta\ket{1}\otimes U_1 U_0 \ket{\psi} ,
\end{split} \end{equation}a coherent superposition of the operators being applied to $\ket{\psi}$ in both possible orderings.
Alternatively, $2\!-switch$ operation could be described by a quantum circuit with three
wires
(shown in Fig.~{\ref{FIG:theoryQpermute}}(a)): one for the control qubit $\ket{c}$, one for the
input
qubit $\ket{\psi}$, and one for an ancilla qubit $\ket{a}$. The input wire leads to the operator combination
$U_0U_1$, while the ancillary wire leads to the reversed operator combination $U_1U_0$. The control qubit
coherently exchanges the states of
the input and ancilla wires via a controlled-swap operator, and the input state $\ket{\psi}$
propagates down a coherent superposition of both the input and ancilla wires, passing through both
sets of operators, before being deterministically returned to the input wire by a second
controlled-swap. In this
way, \textit{the input qubit passes through the quantum operators in both possible orderings}. The circuit depiction requires two separate copies of each box interacting with one input qubit and one
ancilla qubit, whereas the functional description, $2\!-switch\left(U_0,U_1,\ket{c},\ket{\psi} \right) $, enumerates one copy
of each
operator. For a more direct
comparison to the circuit model, the functional description can be illustrated as a circuit diagram
with movable wires. This modified circuit (shown in Fig.~{\ref{FIG:theoryQpermute}}(b)), shows a single copy of $U_0$ and $U_1$ with a superposition of two possible paths through them created by moveable wires. The functional description is less complex, it achieves operator permutation with a single copy of each box and no ancillary qubits.

\paragraph*{}
The reduction in complexity achieved by the functional description of $2\!-switch$ becomes even more significant as the problem is generalized to a permutation of $N$ quantum operators. The functional description, $N\!-switch\left(U_0,U_1\ldots,U_{N-1},\ket{c},\ket{\psi}\right)$,  references each input quantum operator only once while the quantum circuit for implementing an equivalent computation has been shown to require $O(N^2)$ ancillary quantum operators, $O(N^2)$ total extra qubits (including both ancillary and control), and exactly $N$ copies of each quantum operator \cite{D'ArianoProgrammableQuantumComputation}. Although the functional description of $N\!-switch$ is less complex than the circuit description, this does not imply a reduction in \textit{physical} resources. Rather, the quantum functions provide a different, simpler, framework for discussing quantum computations. The value of this is clearly illustrated by its contribution to discussions on the limits of the quantum circuit model and the role of causality in quantum mechanics. Below, we propose an optical device which can implement the $N\!-switch$ using only a single physical copy of each operator.

\section{An Optical Operator Permuting Device}

\paragraph*{} Functional abstraction in classical computing has enabled the development of a myriad of complex programs. Now that a fully quantum-controlled functional model for functional quantum computing has been introduced, it is feasible that analogous development in quantum algorithms may be possible. A proof-of-principle exemplification of this functional model is the $N\!-switch$ operation, which permutes the order in which $N$ input quantum operators are applied to an input qubit register. As shown above, this operation can be visualized as a quantum circuit with movable wires that coherently change the path(s) which input qubits take through the input operators. Implementation of these movable wires may be achieved in an optical device using quantum switches to alter the path of an arbitrarily encoded photonic qubit, or qubit register, such that it encounters optical circuit elements in a particular order. Although the $N$ input operators will act on the physical basis in which the input quantum state is encoded, the order
in which the operators are applied will be coherently controlled via manipulation of the photonic signal's spatial and temporal degrees of freedom (modes).

\begin{figure}[b!]
\includegraphics{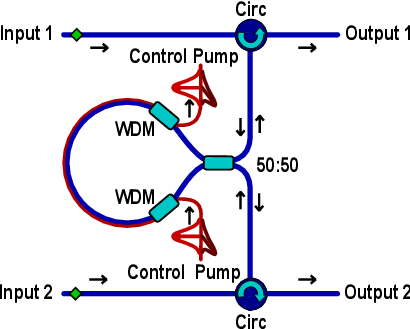}
 \caption{ \label{FIG:opticalSwitch} A schematic for the quantum switch. Cross-phase
modulation, caused by a pump
pulse multiplexed in and out of a fiber loop, alters the quantum signal interference within a
nonlinear optical loop mirror, thereby coherently controlling whether an input signal exits from
output 1 or
output 2.}
\end{figure}

\paragraph*{}
A switch has been demonstrated which uses an optical control pulse to couple the spatial and temporal modes of photonic qubits without otherwise disturbing their quantum state \cite{HallSwitchPRL,HallSwitchNJP,xbarswitch} (see Fig.~{\ref{FIG:opticalSwitch}}). The two-input, two-output, all-optical, fiber-based switch is a modified nonlinear optical loop mirror (NOLM) \cite{NOLM} whose reflectivity is controlled via cross-phase modulation generated by an optical pump (control) pulse. In the absence of a control pulse, a NOLM can be aligned so as to perfectly reflect all incoming signal pulses. This perfect reflection is a result of destructive interference between the clockwise and counter-clockwise paths of the Sagnac loop \cite{loopMirror}. A dual-wavelength control pulse is multiplexed in and out of the clockwise path of the Sagnac loop and consists of two equal-energy cross-polarized optical pulses such that a signal in the clockwise path of the Sagnac loop receives a polarization-independent $\pi$ phase shift \cite{PIXPM}. This phase shift turns destructive interference into constructive interference and causes any signal which passes through the pump pulse to totally transmit through the NOLM. The switch outputs are demultiplexed from the input spatial modes by fiber circulators, giving the device two input and two output spatial modes. Note that although the switching process is governed by pump light in the classical domain, the switch can act on quantum signals without disturbing them because the underlying physical process is coherent. These switches, which couple the spatial and temporal modes of photonic qubits, can be used to implement $N\!-switch$.

The quantum switch can be thought of as having two states: ``on'', where the spatial mode
of an input photonic signal is switched (i.e., an optical control pulse is present), and ``off'', where the
spatial mode of an input photonic signal is not switched (i.e., no control pulse is present). In other
words, the switch performs a conditional bit-flip on a photonic signal's spatial mode. This operation is
conditioned on the presence of a control pulse. Because the photon is a quantum object, a photonic signal can
enter the switch in a coherent superposition of two different temporal modes. If the switch is ``on'' for
one
of those modes and ``off'' for the other, the timing of the photonic signal will coherently control the
conditional spatial bit-flip; i.e., the switching action  acts as a quantum controlled-NOT
operation
\cite{HallSwitchPRL}. The length of the Sagnac loop determines the ``on-time'' of the switch\textemdash the window in which an input photonic signal will be ``switched'' (the smallest demonstrated switching window is 30 ps \cite{xbarswitch}). Additionally, there is some transition time between the off and on
states, approximately equal to the temporal extent of the control pulses (typically 5 ps). Using
these two pieces of timing information, one can define an arbitrary number of orthogonal temporal
modes ($t_0$, $t_1$, etc...), or time-bins, in which a signal will
be switched
contingent on the presence of a control pulse. By constraining a photonic signal to exist in a
coherent superposition of $2^T$ temporal modes \begin{equation}
                                   \sum_{i=0}^{2^T-1}\alpha_i\ket{1}_{t_i}\prod_{j = 1,j \neq
i}^{2^N-1}\ket{0}_{t_j}
                                               \end{equation}where
$\sum^{2^T-1}_{k=0}\vert\alpha_k\vert^2=1$,
 a register of $T$ temporal qubits can be encoded
onto the signal. This can be done by either generating the signal in a superposition of temporal modes as in \cite{Sammy} or by using unbalanced interferometers similar to the one in \cite{TbinGen} to separate the probability amplitude of the signal into multiple time-bins. Additional qubits may be encoded on the photonic signal by using other degrees of freedom such as orthogonal polarization modes, optical frequency modes, photon number, etc... provided the resulting quantum state can be guided by single-mode fiber. Using quantum switches, it is possible to create a device which re-wires the connections between a collection of physical quantum operators that act on these additional qubits. The inter-operator connections can be controlled such that each temporal mode experiences a different series of quantum operations, thereby implementing $N\!-switch$.

\begin{figure*}[htb]
\centerline{\includegraphics{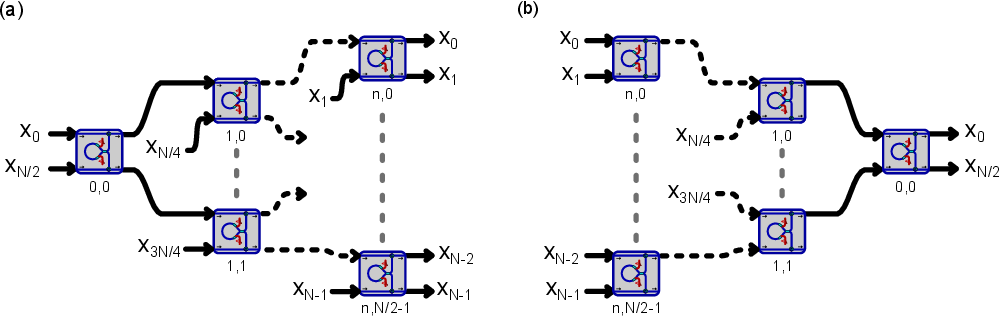}}
 \caption{\label{FIG:mux}(a) A network of $N-1$ quantum switches which multiplexes an input photonic signal from the $x_0$ spatial mode into one of $N$ spatial modes. (b) A network of $N-1$ quantum switches which de-multiplexes $N$ spatial modes of a photonic signal into a single spatial mode ($x_0$).}
\end{figure*}

\paragraph*{}Consider a network of $N - 1$ switches $S_{p,q}$ connected in a binary tree
structure \cite{bTree}, where $p \in \left\lbrace 0,1,...,n=\log_2 N - 1
\right\rbrace$ indicates the level of the tree in which a switch resides and $q \in
\left\lbrace0,1,...,2^{p}-1 \right\rbrace$ is the index of the switch within
it's level (see in Fig.~{\ref{FIG:mux}}(a)). Each level of the tree contains $2^p$ quantum
switches and the tree encompasses a total of $N$ spatial modes ($x_0$, $x_1$, ...,
$x_{N-1}$). Every switch not in the $p=n-1$ level has its outputs connected to input 1 of the two switches in
the next level: $S_{p+1,2q}$ and $S_{p+1,2q+1}$.  For a given $p$ and $q$, a quantum switch acts on
two spatial modes: $x_{ 2qN/2^{p+1}}$  and $x_{(2q+1)N/2^{p+1}}$, where all switches with $p>0$ are
assumed to receive a vacuum field input via the $x_{(2q+1)N/2^{p+1}} $ mode. In an $N=8$ network,
for example, $S_{1,1}$ acts on $x_{2(1)(8)/2^{1+1}}  = x_{ 2(8)/4}  = x_4$ and
$x_{ (2(1)+1)(8)/2^{1+1} }=  x_{3(8)/4}  = x_6 $. Because of the tree
structure, a photonic signal input to the $S_{0,0}$ switch can be
routed to any of $N$ spatial modes at the output. In other words, the tree of quantum
switches is a multiplexer which maps a single input spatial mode to one of
$N$ output spatial modes. If the structure is reversed (see Fig.~{\ref{FIG:mux}}(b)) such that output 1 of both switch $S'_{p+1,2q}$ and switch $S'_{p+1,2q+1}$ connect to input 1 and input 2 of switch $S'_{p,q}$, the switch network becomes a demultiplexer capable of routing a photonic input from any input spatial mode of a switch in the $p = \log_{2}(N) - 1$ level to either output spatial mode of the
$S'_{0,0}$ quantum switch. In a de-multiplexer switch network, all
switches with $p>0$ are assumed to output a vacuum field via the $x_{
(2q+1)N/2^{p+1}}$ mode. To avoid confusion, $S'$ will refer to switches used for
demultiplexing and $S$ to switches for multiplexing.

\begin{figure*}[htb]
\centerline{\includegraphics{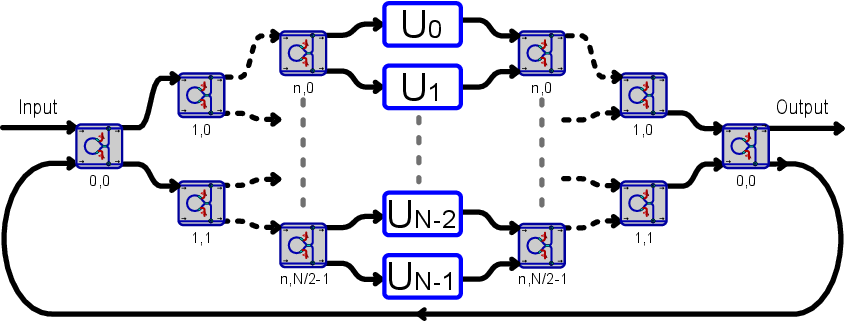}}
 \caption{ \label{FIG:opticalDevice}  An optical device which simulates $N\!-switch$
using $2N-2$
quantum switches, and $N$ physical quantum operators. Because the
behavior of the quantum switches is governed by a predetermined pattern of optical control
pulses, the time at which
a photonic signal is input to the device can be mapped to an order in which the photon
passes through the quantum operators.}
\end{figure*}

\paragraph*{}Combining a multiplexer quantum switch network whose output
spatial modes are all connected to different physical quantum operators
($x_0\!\rightarrow\! U_0, x_1\!\rightarrow \! U_1,..., x_{N-1}\!\rightarrow \! U_{N-1}$), with a
demultiplexer quantum switch network, whose inputs are all
connected to the outputs of the quantum operators, yields a device which can route an input signal
through any of $N$ unitary operators and return it to its initial spatial mode (see
Fig.~{\ref{FIG:opticalDevice}}). With the
addition of a return loop connecting output 2 of $S'_{0,0}$ to input 2 of $S_{0,0}$ via $x_{N/2}$, the photonic signal can be routed through the device an arbitrary number of times, $M$, passing through an arbitrary quantum operator in each iteration. In other words, the device could take an input photonic signal via the $x_0$ input of $S_{0,0}$, apply one of $N^M$ possible combinations of quantum operators, and output the result to $x_{0}$ via output 1 of $S'_{0,0}$. By programming the device to route the photonic signal through a different combination of operators depending on its temporal mode, the functionality of the device can be significantly expanded.

\paragraph*{}The network of quantum switches can be programmed such that a photonic signal is routed through a unique permutation of $N$ quantum operators for each of it's temporal modes in $M=N$ iterations through the device. In this scheme, the state of an input photonic signal is
\begin{equation}
 \sum_{i=0}^{N^N-1}\alpha_i \ket{\psi}_{t_i , x_0} \prod_{j=0,j\neq i}^{N^N-1} \ket{0}_{t_j , x_0}
\end{equation} and the temporal modes are serialized by number and chosen such that $t_{N^N-1}$
passes through the input switch before $t_0$ enters the return loop for the first time; the output
state is \begin{equation}
 \sum_{i=0}^{N^N-1}\alpha_i O_i \ket{\psi}_{t_i , x_0} \prod_{j=0,j\neq i}^{N^N-1} \ket{0}_{t_j ,
x_0},
\end{equation} where each temporal mode ($t_i$) has undergone a different combination of
quantum operators $O_i = O_{i,N-1}O_{i,N-2}...O_{i,0}$, and $O_{i,k}= U_{l}$ where
\begin{equation} l = \left\lfloor\frac{i}{N^{N-j-1}} \right\rfloor \text{mod} \: N, \end{equation}
and $j \in \lbrace{1,2,\ldots,N-1\rbrace}$ indexes each iteration of the photonic signal through the device. 

\paragraph*{} Each $U_{l}$ can be applied to an input photonic signal by activating a unique group of switches such
that they are in the ``on'' state while interacting with the $i$th temporal mode of a signal and in the
off state, if necessary, before the next temporal
mode of the signal will interact with them. To apply the $U_l$ operator to a specific temporal mode of the
signal, all switches with indices such that \begin{equation}   \left( q+1\right)
\frac{N}{2^{p+1}} \leq l < \left(q+2 \right)\frac{N}{2^{p+1}}, \end{equation}
and $p\neq0$ must be turned on when that temporal mode enters the switch. Because
the $p=0$ switches also serve as input/output ports to the device, their behavior must be different. The $S_{0,0}$ switch must be turned on if either: 1) $j=0$ and $l \geq \frac{N}{2}$, or 2) $j\neq0$ and $l < \frac{N}{2}$. The $S'_{0,0}$ switch must be turned on if either: 1)
both $j=M-1$ and $l\geq \frac{N}{2}$, or 2) $j\neq 0$ and $l<\frac{N}{2}$. By
applying these switching rules for every iteration of every
temporal mode of the input signal, a superposition of all $N^N$ possible operator permutations is applied
to the input photonic quantum state $\ket{\psi}$. Note that this device is more general than an $N\!-switch$ because it can be used to control not just the ordering, but also number of uses of each operator. This device will deterministically implement the
$N\!-switch$ operation if the input signal is superposed over only those temporal modes to which non-degenerate operator permutations will be applied. 

\paragraph*{}
The savvy reader may note that our description of switching rules means that the required number of temporal modes scales as $N^N$, and can become unreasonable very quickly. However, the required number of modes doesn't exceed 1000 until one considers cases where $N\! >\! 4$. We suggest that our method be considered for small-scale experiments, as there are no existing proposals for an implementation of $N\!-switch$, $N\! >\! 2$,which is both deterministic and scalable. We choose 1000 time-bins as the cut-off number for a ``reasonable'' experiment because with a 1-GHZ system clock 1000 time-bins would span 1$\mu $s in time and $\sim $200 m in space. Truly, there are only $N!$ operator permutations where no operator is called more than once, so the number of time-bins could be reduced accordingly in this case. If the system is reconfigured such that it only processes $N!$ temporal modes, then the required number of modes doesn't exceed 1000 unless $N\! >\! 7$.

\paragraph*{}
One application of optical operator permutation is that it may be used to measure the commutativity of two physical operators. Consider that the state output from an optical $2\!-switch\left( U_0,U_1, \frac{1}{\sqrt{2}}\left(\ket{0}+\ket{1} \right), \ket{\psi}\right)$ device is \begin{equation}
\frac{1}{\sqrt(2)}U_1U_0\ket{\psi}_{t_0 x_0} + \frac{1}{\sqrt(2)}U_0U_1\ket{\psi}_{t_1 x_0},
\end{equation} where modes containing vacuum are left out for simplicity. Using an additional switch, the amplitude in $t_0 x_0$ can be routed into an ancillary spatial mode $x_a$ as shown in Fig. 5. The spatially separated signal amplitudes can then be interfered by placing an optical delay (i.e. a short length of fiber) in $x_a$ equal to the time between temporal modes and combining $t_1 x_0$ and $t_0 x_a$ on a 50:50 beamsplitter. This interference will yield the state \begin{equation}
\frac{1}{2}\left(U_1U_0 - U_0U_1 \right)\ket{\psi}_{t_1 x_0} + \frac{1}{2}\left(U_1 U_0 + U_0 U_1 \right)\ket{\psi}_{t_1 x_a}. 
\end{equation}
The commutativity of the two operators, $\left[U_0,U_1\right] = U_1U_0 - U_0U_1$, can be determined by post-selectively detecting only the signal in $x_0$, and using quantum state tomography to reconstruct the state prepared in that spatial mode. 

\paragraph*{} It should be noted that \cite{ExperimentalSuperposition} reports an experimental measurement of operator commutativity, but using \textit{spatial} modes to control operator ordering. This is a simpler approach for performing a $2\!-switch$, because the control information is encoded using a single beamsplitter and the experiment contains only passive components. However, it relies upon multiple spatially distinct modes propagating through the same set of free-space optical components which poses obvious issues for scalability. 

Another work which does address the permutation of an orbitrary number of operators is \cite{ComputationalAdvantage}. This is primarily a theoretical treatment which introduces the concept of an $N\!-router$ that applies a variable re-mapping of the target quantum state's modes such that each mode can be arbitrarily coupled to one of the input operators. In this work, another $N\!-router$ would then apply the inverse mode transformation and output the result back to an input mode of the first $N\!-router$, repeating as necessary to implement the full operator permutation. Clearly, our multiplexer and de-multiplexer switch networks are $N\!-routers$ which apply a time-to-space mode mapping. In contrast to our work, the authors of \cite{ComputationalAdvantage} suggest that the that the control qubits could be encoded over orbital angular momentum (OAM) states and that OAM mode sorters could be used to couple different modes to different operators. This approach has the advantage that the dimensionality of the target qubit will not be expanded in the time, a factor which could ease practical constraints on the time-scales of system stability and lead to a significant reduction in computation time compared to our proposal. The disadvantage to this approach is that current state-of-the-art OAM mode sorters have slower update rates ($\sim$4 kHz) and lower extinction (93$\%$)\cite{BoydOAM} than the switching technology which can update as fast as 10 GHz\citep{xbarswitch} with $\> 99\%$ mode extinction\cite{HallSwitchPRL,HallSwitchNJP}.

\begin{figure*}[htb]
\centerline{
\includegraphics{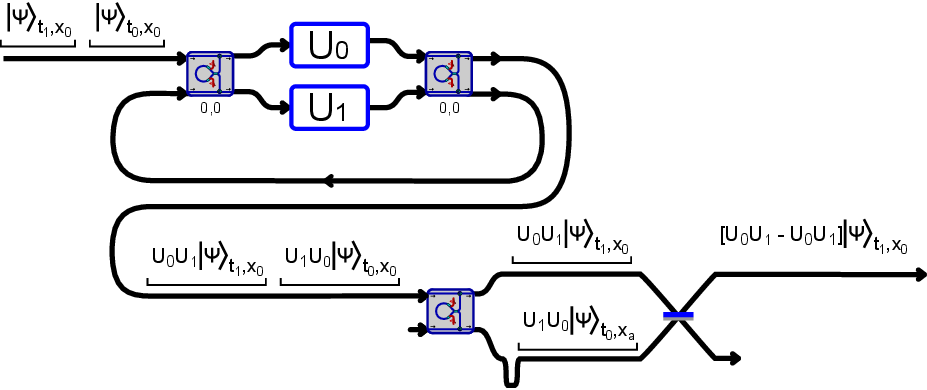}
}
\label{FIG:commut}
\caption{Experimental design for measuring the commutativity of two operators}
\end{figure*}

\paragraph*{}One potential experimental issue with our approach is that a given quantum operator will interact with each temporal
mode of the input photon at a different time. At each of these different times,
the transformation applied by the operator to the photonic signal's quantum state may be varied
depending on air turbulence, vibration of optical components, noise in electrical components, etc\ldots In
essence, the proposed device does not apply the quantum operator $N$ times.
Instead, it applies $N$ different approximations of the operator. One approach to overcoming this
problem, if necessary, is to exploit the Fourier relationship between time and frequency. Applying a narrowband
frequency filter to the photon at the device output will broaden its temporal profile, though such
filtering could be a significant source of loss.
If sufficiently narrow filtering is possible without attenuating the signal below
measurable levels, it may become impossible (even in principle) to
determine whether or not the different temporal modes of the photon interacted with different
approximations of each operator. In this case, every temporal mode of the photon has effectively
interacted with the same approximation of each operator, and the $N\!-switch$
operation has
been implemented.

\section{Conclusion}
\paragraph*{}

A new model of functional quantum computation has recently been presented, one which describes quantum computers as a collection of quantum operators with programmable interconnections. Within this model, a proof-of-principle operation, $N\!-switch$, has been proposed which takes as input: $N$ quantum operators, a quantum control register, and an input quantum state; the input operators are coherently re-ordered according to the state of the control register and then applied to the input state. While operator permutation is currently of interest both fundamental science and quantum information processing, there has been no discussion of practical methods to permute more than two operators. We have proposed an all-optical device to implement $N\!-switch$ for a photonic signal whose superposition over multiple temporal modes encodes the control qubits for the permutation and whose other degrees of freedom are used to encode the computational quantum state. This all-optical device utilizes $2N-2$ quantum switches, and a single physical implementation of each quantum operator to be permuted. We have also presented a scheme in which an optical $N\!-switch$ device may be used to test the commutativity of two physical quantum operators. Although the optical $N\!-switch$ device is not scale-able, it could be used for small or test-bed systems. 

This work was supported in part by the National Aeronautics and Space Administration (NASA) Space Technology Research Fellowship (Grant No. NNX12AN27H). Although this work has been available as arXiv:1211.1257v1 since 2012, it has only now been submitted for peer review due to a professional conflict for PK.
\bibliography{fqc}
\end{document}